# The LOFT Ground Segment

E. Bozzo[*a], A. Antonelli[b], A. Argan[c], D. Barret[d,r], P. Binko[a], S. Brandt[e], E. Cavazzuti[b], T. Courvoisier[a], J.W. den Herder[f], M. Feroci[g,q], C. Ferrigno[a], P. Giommi[b], D. Götz[h], L. Guy[a], M. Hernanz[i], J.J.M. in't Zand[f], D. Klochkov[l], E. Kuulkers[j], C. Motch[k], D. Lumb[p], A. Papitto[i], C. Pittori[b], R. Rohlfs[a], A. Santangelo[l], C. Schmid[m], A.D. Schwope[n], P.J. Smith[o], N. A. Webb[d,r], J. Wilms[m], S. Zane[o]

[a]ISDC, University of Geneva, Chemin d'Ecogia 16, 1290, Versoix, Switzerland; [b]ASDC, Via del Politecnico snc - 00133 Rome, Italy; [c]INAF HQ, Viale del Parco Mellini 84, 00136, Rome, Italy; [d]CNRS, IRAP, 9 Avenue du Colonel Roche, BP44346, 31028, Toulouse, France; [e]National Space Institute, Technical University of Denmark, Elektrovej Bld 327, 2800 Kgs Lyngby, Denmark; [f]SRON, The Netherlands Institute of Space Research, Utrecht, The Netherlands; [g]INAF-IAPS-Roma via Fosso del Cavaliere, 100, 00133, Rome, Italy; [h]CEA Saclay, DSM/IRFU/SAp, 91191 Gif sur Yvette, France; [i]IEEC/CSIC, Campus UAB, E-08193, Bellaterra, Spain; [j]European Space Astronomy Centre (ESA/ESAC), Science Operations Department, 28691 Villanueva de la Cañada, Madrid, Spain; [k]Observatoire Astronomique de Strasbourg, 11 rue de l'Université - 67000 Strasbourg, France; [l]IAAT University of Tuebingen, Sand 1 - 72076 Tuebingen, Germany; [m]University of Erlangen-Nuremberg, Schlossplatz 4 - 91054 Erlangen, Germany; [n]Leibniz-Institut fuer Astrophysik Potsdam, An der Sternwarte 16 - 14482 Potsdam, Germany; [o]Mullard Space Science Laboratory, UCL, Holmbury St Mary, Dorking, Surrey, RH56NT,UK; [p]Astrophysics & Fundamental Physics Mission Division, ESA/ESTEC, Noordwijk, Netherlands; [q]INFN, Sez. Roma Tor Vergata, Via della Ricerca Scientifica 1 - 00133 Rome, Italy ; [r]Universite de Toulouse; UPS-OMP; IRAP; Toulouse, France

on Behalf of the LOFT Consortium

**ABSTRACT**

LOFT, the Large Observatory For X-ray Timing, was one of the ESA M3 mission candidates that completed their assessment phase at the end of 2013. LOFT is equipped with two instruments, the Large Area Detector (LAD) and the Wide Field Monitor (WFM). The LAD performs pointed observations of several targets per orbit (~90 minutes), providing roughly ~80 GB of proprietary data per day (the proprietary period will be 12 months). The WFM continuously monitors about 1/3 of the sky at a time and provides data for about ~100 sources a day, resulting in a total of ~20 GB of additional telemetry. The LOFT Burst alert System additionally identifies on-board bright impulsive events (e.g., Gamma-ray Bursts, GRBs) and broadcasts the corresponding position and trigger time to the ground using a dedicated system of ~15 VHF receivers. All WFM data are planned to be made public immediately. In this contribution we summarize the planned organization of the LOFT ground segment (GS), as established in the mission Yellow Book[1]. We describe the expected GS contributions from ESA and the LOFT consortium. A review is provided of the planned LOFT data products and the details of the data flow, archiving and distribution. Despite LOFT was not selected for launch within the M3 call, its long assessment phase (> 2 years) led to a very solid mission design and an efficient planning of its ground operations.

---

# 1. INTRODUCTION

LOFT, the Large Observatory For X-ray Timing, is one of the M3 candidate missions that completed at the end of 2013 the assessment phase and competed for a launch of opportunity in 2022-2024. LOFT was designed to answer fundamental questions about the motion of matter orbiting close to the event horizon of a black hole, and the state of matter in neutron stars [1]. The LOFT payload comprises:

- The Large Area Detector (LAD): a collimated experiment performing pointed observations of X-ray sources mainly in the 2-30 keV energy range. Its unprecedentedly large collecting area for X-ray photons, reaching 10 m$^2$ at 8 keV, is enabled by the usage of 2016 innovative Silicon Drift Detectors and low-weight micro-channel plate collimators [1]. The LAD will provide about 240,000 cts/s for a 1 Crab target, achieving an energy resolution of <240 eV (down to ~180 eV for 40% of the events) and a time resolution of 10 μs.

- The Wide Field Monitor (WFM) is a coded mask instrument [3], comprising 10 identical cameras each equipped with its own coded mask and 4 SDDs (similar to those employed for the LAD). Each camera has a FoV of 90°x90° and the whole instrument is able to monitor more than 1/3 of the sky at once. The WFM performs observations around the direction of the LAD pointing in the 2-50 keV energy range and achieves an energy (time) resolution of 300 eV (10 μs). This instrument is also equipped with an on-board system to detect bright impulsive astrophysical events, such as GRBs, magnetar flares, etc [4]. The triggers are broadcast to the ground within 30 s of the identification in the form of short messages through a dedicated system of about 15 VHF receivers. The latter are deployed around the Earth's equator to cover the low equatorial orbit of the satellite.

LOFT is planned to be operated as an observatory open to the community, with observing programs executed in response to yearly "Announcements of Opportunities". LAD programs require durations ranging from a few to hundreds of ks, including standard, fixed time, monitoring programs and target of opportunity (ToO) observations. The science drivers of LOFT are detectable in several persistent sources (e.g., the relativistic broadened Iron lines in AGNs) but they are expected at best in extreme states of Galactic sources, like the outbursts of black hole transients. LOFT is also expected to devote a large fraction of its total observational time to secondary science goals ("observatory science"). These include the study of X-ray emissions from a wide range of celestial objects and the coordination of "multi-messenger" observational campaigns with other facilities that could be operating at the time of LOFT.

To fulfil the main science goals of LOFT and to best exploit all achievable observatory science, mission operations should be capable of reacting to triggers based on the Wide Field Monitor data and/or external triggers from other multi-frequency observatories. The LOFT ground segment (GS) has thus been designed with particular emphasis on the following:

- It is important that the LOFT mission identify relevant transient events and state changes in the observable sources. These identifications will mostly be done by performing a quick-look processing of the WFM and LAD data by the ground segment, but can in principle also be provided through other means (i.e. other facilities operating at the same time as LOFT).
- For a significant fraction of the targets an observational campaign can be initiated following the triggers on the source that are provided by the above processing and analysis. These campaigns consist of repeated observations of a source of up to several months [1]. In order not to spend observing time on sources which are either too weak or not interesting, sources should be monitored through the Quick-Look processing of the WFM/LAD data to decide whether to continue or to interrupt the observation of a source.
- The alerts delivered by the LOFT Burst Alert System (LBAS) might or might not correspond to events related to the main LOFT science goals but their prompt distribution to the science community at large can significantly improve the science return of the mission.

## 1.1 The ground segment overview

An overview of the planned LOFT GS organization is provided in Figure 1. This is a combined effort of ESA and the LOFT consortium. The division of responsibilities between ESA and the consortium was established during the mission assessment phase in order to optimize the use of available resources and the functionalities of the GS.

The LOFT GS comprises the Operational Ground Segment (OGS) and the Science Ground Segment (SGS). The OGS includes the main ground stations for the telemetry reception and commands uplink and the Mission Operation Center

(MOC). Due to the large throughput of the LOFT mission, it is foreseen to use a single ground station for the satellite commanding and two ground stations for the telemetry downlink. The SGS includes the Science Operation Center (SOC), the Instruments Operation Centers (IOCs), the Science Data Center (SDC) and the LOFT Burst Alert Ground Segment (LBAGS). The LBAGS comprises a network of VHF stations, which receive the alerts from the on-board VHF transmitter, and the LOFT Alert Center (LAC), that is in charge of validating these alerts and broadcasting further relevant updates on to the science community. The SDC, the IOCs, and the LBAGS are provided by the LOFT consortium.

## 1.2 Data products

Before describing the tasks of the different GS parties, we summarize below the different types of data products defined so far for the LOFT mission (see also Table 1):

- *Raw telemetry*: this includes instrument science telemetry as well as preliminary auxiliary data as broadcast from the spacecraft to the ground stations.

- *Level 0*: raw spacecraft telemetry is de-commutated and split into functionally independent parallel streams formatted as binary FITS files. The data is organized per observation and readable with standard tools. On-board calibrations are applied to both the LAD and WFM data.

- *Level 1*: all corrections (such as aspect correction, time calibration, barycentric corrections) and instrument-specific calibrations (such as detector gain and good-timing information) are applied to level 0 data to produce cleaned FITS event files for both the WFM and LAD.

- *Level 2*: data level 1 from multiple observation intervals that constitute an observation are combined to create uniform sets of products for the WFM and LAD.

- *Level 3*: enhanced, higher level, scientific products derived from Level 2 data. These comprise, e.g., catalogues, mosaics, and products like long-term light-curves of individual sources.

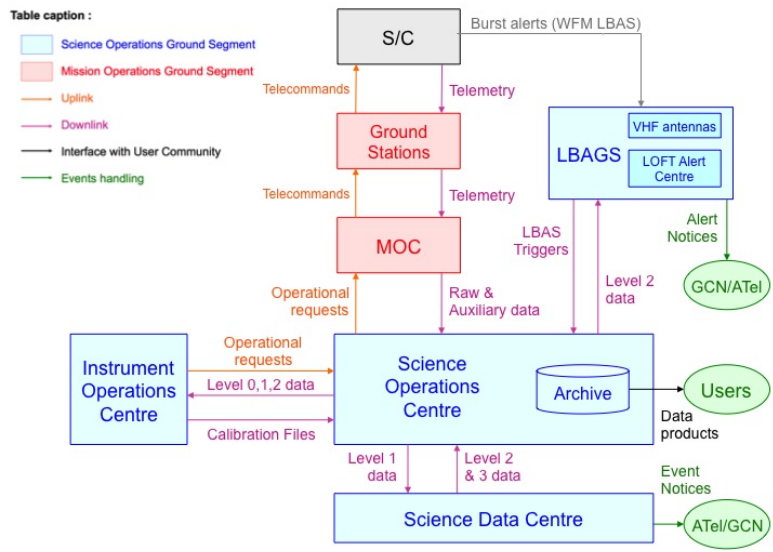

Figure 1 Overall organization of the LOFT GS (figure from the LOFT Science Operation Assumption Document, SOAD)

## 2. THE GROUND SEGMENT ORGANIZATION

### 2.1 The Mission operation center and the Science operation center

The MOC and SOC cover for LOFT standard tasks, similar to those provided for a number of previously operated X-ray observatories. They are under ESA responsibility.

MOC tasks comprise the following: remote command of the spacecraft and instruments (including overall mission planning), assurance of the spacecraft safety and health, provision of flight dynamics support (including determination and control of the satellite's orbit and attitude), and intervention in the case of anomalies. The MOC performs all communications with the satellite through the ground stations for the upload of the platform and payload commands (based on the observation schedule provided by the SOC), and reception of the downloaded telemetry data. MOC is also responsible for collecting the science data and their transmission to the SOC (along with the raw telemetry, housekeeping and auxiliary data). The SOC is the unique point of contact to the MOC for providing detailed operational requests, and plans the payload operation activities.

The SOC is also the first point of contact of the scientific user community, providing a centralized helpdesk. It handles the observational proposals and requests for ToOs, preparing the corresponding plan for science observations. Specifically in the case of LOFT, the SOC performs also automatically *in situ* processing of level 0 data into level 1 data, and hosts the LOFT Science Data Archive (LSDA). The latter contains all data, calibrations, software, auxiliary files, and science products related to the LOFT mission. The automated quick-look software alerts the SOC about relevant variability and/or state transitions of LOFT core science targets (see §3), in such a way that the relevant science Principal Investigators (PIs) can be promptly alerted by the SOC and a ToO can eventually be planned.

Table 1: Overview of the LOFT data products. Yellow boxes indicate products for which a proprietary period of 1 yr applies. Items in blue boxes are made publicly available as soon as they are ingested in the LSDA. The concept of "Near Real Time" (NRT) and "Consolidated" (CONS) data is clarified in §3.

|  | LAD | WFM | NRT | CONS |
|---|---|---|---|---|
| Level 0 | - FITS event files (preliminary on-board calibrations applied) | - Sky images (sliced in time and energy)<br>- Rate meters<br>- Photon-by-photon (for LBAS triggers) | Y | Y |
| Level 1 | - Cleaned, corrected and barycentered event files (all ground calibrations and corrections applied) | - Same as above but with all corrections and ground calibrations applied) | Y | Y |
| Level 2 | For the target source:<br>- Lightcurve<br>- Energy spectrum<br>- Power spectrum | - Sky mosaic<br>For all sources in the FOV:<br>- Lightcurve<br>- Energy spectrum<br>- Power spectrum (if applicable) | Y | Y |
| Level 3 | - Multi-wavelength spectral energy distribution | - All sky mosaic (continuously updated)<br>- Historical database of fluxes and spectral/timing states for all sources)<br>- Cross-catalogue identification | N | Y |
| VHF Alerts | - | - Short text messages containing start time and position (1 arcmin accuracy) of LBAS triggers (fulfilling pre-defined criteria) | - | - |

## 2.2 The Science Data Center

The SDC is planned to be managed as a consortium of different institutes, which are responsible for developing and providing all software needed to process, analyze, and manipulate LOFT data. The SDC is also in charge of performing automatic processing of the level 1 data into level 2 and level 3 data by making use of a preferential access to the LSDA. All the derived science products are stored in the LSDA. The SDC is responsible for the scientific validation of all LOFT data and for providing support to the community for everything concerning data products, analysis and software issues (through the centralized help desk at SOC). The IOCs support the SDC in these activities.

As part of the SDC tasks, it is also foreseen to carry out a "sky monitoring" using the WFM data. All sources observed by the WFM are monitored by automatically extracting a quick-look version of the level 2 data. These products are made publicly accessible in near real time to the science community at large, in order to maximize the usage of data and the science return of LOFT (see §3).

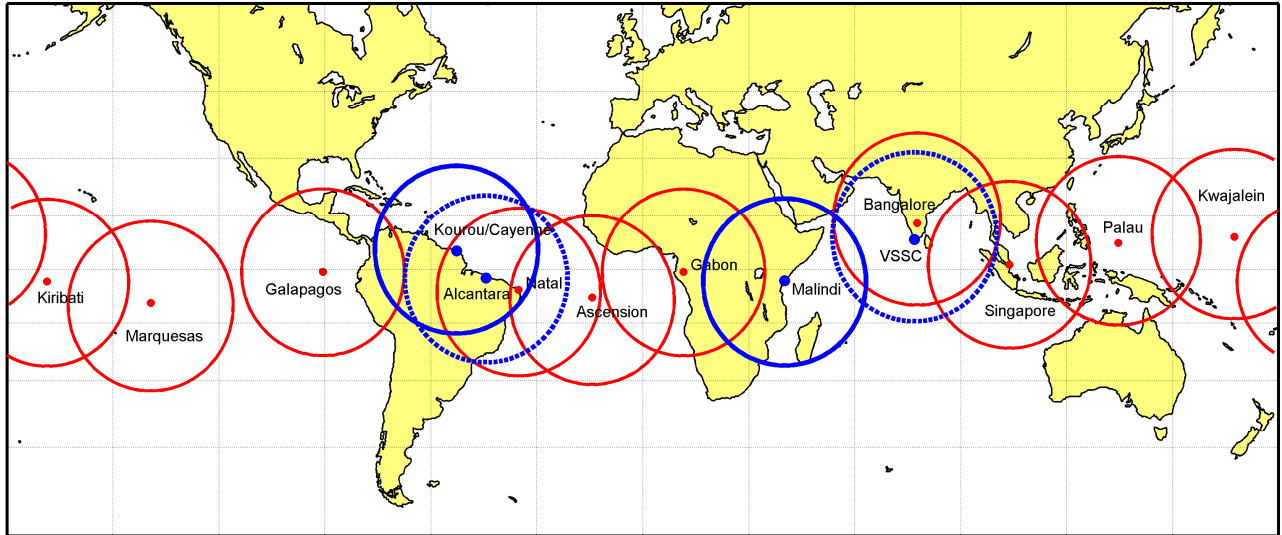

Figure 2 Locations of different ground stations to be used for LOFT. Possible candidate antennae for the main telemetry downlink and commands uplink are represented in blue; the LBAGS VHF receivers are marked in red (figure from the LOFT Yellow Book).

### 2.3 The Instruments Operation Centers

In the LOFT GS two IOCs are foreseen. These comprise part of the LAD and WFM instrument teams residing in different countries of the LOFT consortium. The IOCs provide support to ESA for maintenance of the instruments after launch, operations, processing algorithms and calibration. They will be responsible for characterizing and calibrating the instrument responses, for monitoring the science performances, and carrying out their long-term trend analysis.
They provide inputs for the instrument specific software to the SDC. The IOCs also assist personnel at SOC to acquire the necessary knowledge on LOFT instruments and secure their long-term maintenance.

### 2.4 The LOFT Burst Alert Ground Segment

The LBAGS has been designed in order to optimize the dissemination of alerts provided by the LBAS on-board LOFT. The LBAS will identify bright impulsive astrophysical events and send the position (~1 arcmin accuracy) and the time of the event in the form of short messages to the ground through an on-board VHF transmitter.
Each alert is received by one or more of the VHF antennae deployed along the Earth's equator (see Figure 2). The VHF receivers cover an as large as possible part of the LOFT orbit and relay the information through the internet to the LOFT Alert Centre (LAC) and other users that have registered with the service. It is estimated that, after initial verification, the time between the onboard detection of the transient events to the end user will be less than 30 s.
The WFM data corresponding to the trigger (event-by-event data and images) are telemetered to the MOC and processed at the SOC (raw to Level 0 to Level 1) and SDC (Level 1 to Level 2); during normal conditions the full instrument data are publicly available no later than 3 hours after the event. The LAC uses these WFM science data to perform the quality checks on the alerts, and receives from the WFM IOC all information regarding the health of the WFM and possible hardware changes that might affect the reliability of the triggers. Relevant information are disseminated from the LAC to the science community.

## 3. DATA FLOW, POLICY AND PROCESSING STRATEGY

The overall data flow of the LOFT mission is indicated in Figure 1, and a closer look at the processing strategy is shown in Figure 3. In the current organization of the LOFT GS, the raw telemetry is downlinked to the ground from the spacecraft through the two main LOFT ground stations and collected at the MOC. The preliminary WFM telemetry is available to the MOC within 2 satellite orbits, while the LAD telemetry takes up to 1 day (the difference between the two instruments is due to the required reliability in data transmission).
The preliminary telemetry is continuously provided from the MOC to the SOC, where it is automatically processed into Near Real Time (NRT) level 0 and level 1 data. For the NRT processing, the required preliminary attitude and time correlation data are derived from the on-board GPS system and information from the star trackers (made available together with the NRT WFM telemetry). The motivation for having NRT data is to perform a timely quick-look analysis that leads to the identification of interesting events to be followed-up by a dedicated LAD program.

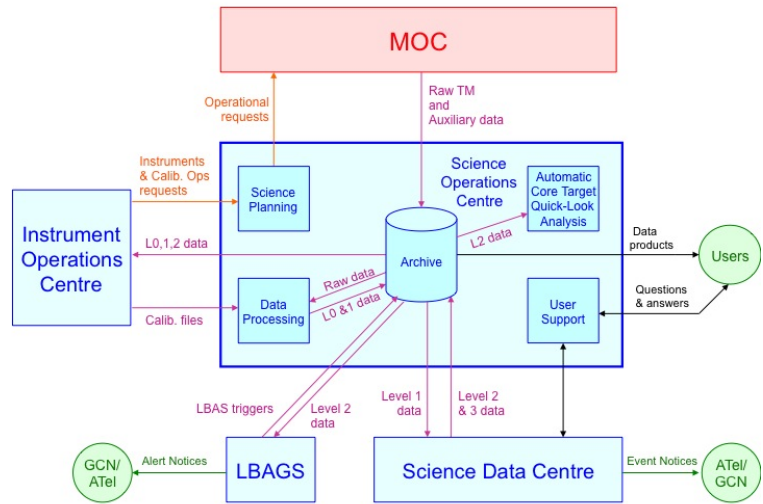

Figure 3 The LOFT data processing strategy (figure from the LOFT SOAD).

In order to correct for problems that might occur in the NRT data transmission, the telemetry is consolidated at the MOC and provided again to the SOC within ~1 week. Refined auxiliary data (including spacecraft flight-dynamics data and other ancillary data) are also provided by the MOC to the SOC on the same timescale. The consolidated telemetry and refined auxiliary data are reprocessed to obtain the final CONS Level 0 and level 1 data. These correspond to the most complete data set that can be obtained and should be used for final scientific analyses. Level 0 is the lowest data level accessible to the community, as scientists may want to reprocess level 0 into level 1 data at any time to apply updated on-ground calibrations.

The SDC makes use of preferential access to the LSDA (granted by the SOC) in order to automatically process available NRT and CONS level 1 data into NRT and CONS level 2 data. The WFM level 2 NRT and CONS data are made public as soon as available, in order to maximize the exploitation of these products by the science community at large.

The SDC routinely monitors the WFM NRT level 2 data. As a result of this "sky monitoring" activity, announcements can be communicated to the science community through, e.g., Astronomer Telegrams and/or GCN circulars.

The quick-look analysis (QLA) software provides the means to inspect and visualize all level 2 data. Customized level 2 data can be produced by any user from level 1 data using interactive tools. The QLA software installed at SOC provides automated alerts in the event that core LOFT science targets undergo interesting state changes and guarantees fast access to LAD NRT data. The latter are used only by the SOC to communicate with the PIs and inform them about the on-going observation and to plan further pointings of the corresponding target. LAD NRT and CONS data are distributed to the PIs through the LSDA. They become publicly accessible when the proprietary period expires (12 months).

Level 3 (enhanced) data are produced only from publicly available CONS WFM and LAD level 2 data. Some of the level 3 data are also stored and made accessible to the science community through the LSDA to provide legacy products.

## 4. CONCLUSIONS

The concept of the LOFT GS presented here has been designed on the heritage of previous X-ray science missions and optimized to fulfil the LOFT science goals. It envisages a certain number of relatively standard tasks for the different parties involved, but provides also a flexible and efficient system for a quick-look analysis of all data and to broadcast alert messages from the on-board LBAS. The nature of the LOFT mission requires, indeed, a dedicated effort to identify relevant transient sources and their state changes. The QLA of the NRT data will provide the means to identify and eventually follow-up the events requiring the fastest reactions; CONS data can be used to investigate long term variability of all sources. As the WFM will observe more than 1/3 of the sky at once in a broad-band energy range with good time and spectral resolution, all NRT and CONS data are immediately made publicly available in order to maximize the involvement of the community and the science return of the mission. LAD data will remain proprietary for one year. LOFT was not selected for launch within the M3 call, but its long assessment phase (> 2 years) led to a solid mission design. The LOFT ground segment concept discussed here has been presented as part of that design and described to some extend in the mission Yellow Book.

## ACKNOWLEDGEMENT

The work of the MSSL-UCL is supported by the UK Space Agency. The work of SRON is funded by the Dutch national science foundation (NWO). The group at the University of Geneva is supported by the Swiss Space Office. The Italian team is grateful for support by ASI (under contract I/021/12/0-186/12), INAF and INFN. The work of IAAT on LOFT is supported by Germany's national research center for aeronautics and space (DRL). The work of the IRAP group is supported by the French Space Agency (CNES). LOFT work at ECAP is supported by DLR (under grant number 50 OO 1111). IEEC-CSIC has been supported by the Spanish MINECO (under grant AYA2011-24704).